\documentclass[showpacs,prb,twocolumn]{revtex4}
\usepackage[dvipdfm]{graphicx}
\usepackage{dcolumn}
\usepackage{bm}
\usepackage{color}
\usepackage{ulem}

\begin{document}
\newcommand{\avrg}[1]{\left\langle #1 \right\rangle}
\newcommand{\eqsa}[1]{\begin{eqnarray} #1 \end{eqnarray}}
\newcommand{\eqwd}[1]{\begin{widetext}\begin{eqnarray} #1 \end{eqnarray}\end{widetext}}
\newcommand{\hatd}[2]{\hat{ #1 }^{\dagger}_{ #2 }}
\newcommand{\hatn}[2]{\hat{ #1 }^{\ }_{ #2 }}
\newcommand{\wdtd}[2]{\widetilde{ #1 }^{\dagger}_{ #2 }}
\newcommand{\wdtn}[2]{\widetilde{ #1 }^{\ }_{ #2 }}
\newcommand{\cond}[1]{\overline{ #1 }_{0}}
\newcommand{\conp}[2]{\overline{ #1 }_{0#2}}
\newcommand{\nn}{\nonumber\\}
\newcommand{\cdt}{$\cdot$}
\newcommand{\bra}[1]{\langle#1|}
\newcommand{\ket}[1]{|#1\rangle}
\newcommand{\braket}[2]{\langle #1 | #2 \rangle}
\newcommand{\bvec}[1]{\mbox{\boldmath$#1$}}
\newcommand{\blue}[1]{{#1}}
\newcommand{\bl}[1]{{#1}}
\newcommand{\bn}[1]{\textcolor{blue}{#1}}
\newcommand{\rr}[1]{{#1}}
\newcommand{\bu}[1]{\textcolor{blue}{#1}}
\newcommand{\red}[1]{{#1}}
\newcommand{\fj}[1]{{#1}}
\newcommand{\green}[1]{{#1}}
\newcommand{\gr}[1]{\textcolor{green}{#1}}
\definecolor{green}{rgb}{0,0.5,0.1}
\definecolor{blue}{rgb}{0,0,0.8}
\preprint{APS/123-QED}

\title{
Composite-Fermion
Theory for Pseudogap,
Fermi Arc,
Hole Pocket, and Non-Fermi-Liquid
of Underdoped Cuprate Superconductors
}
\author{Youhei Yamaji}
\email{yamaji@solis.t.u-tokyo.ac.jp}
\author{Masatoshi Imada}
\affiliation{Department of Applied Physics, University of Tokyo, Hongo, Bunkyo-ku, Tokyo, 113-8656, Japan.}%
\date{\today}

\begin{abstract}
We propose that an extension of the exciton concept to doped Mott insulators offers a fruitful insight into challenging issues of the 
copper oxide
superconductors.   
In our extension, new fermionic excitations called cofermions emerge in conjunction to generalized excitons.
The cofermions hybridize with conventional quasiparticles.
Then a hybridization gap opens,
and is identified as the 
pseudogap observed in the underdoped cuprates.
The resultant Fermi-surface reconstruction naturally explains
a number of unusual properties of 
the underdoped cuprates, 
such as the Fermi arc and/or pocket formation.
\end{abstract}

\maketitle

Since the discovery of cuprate superconductors,
the nature of low-energy electronic excitations evolving
in their normal metallic phase 
has attracted much attention as one of the central issues 
in condensed matter physics.
One reason for the interest lies in its connection to
the origin of the high temperature
superconductivity itself.

Electronic states in the underdoped
cuprates are 
unconventional.
For example, spin and charge excitations are unexpectedly suppressed
as ``pseudogap phenomena" in the normal state.
Recent improvement of experimental tools, such as
angle-resolved photoemission spectroscopy (ARPES),
has further enabled 
resolving
strong momentum dependence of
quasiparticles\cite{Damascelli_RMP,Yoshida06,Meng}.
In particular,
quasiparticles
are hardly observed around
antinodal points
$(\pi,0)$ and $(0,\pi)$ in the 2D Brillouin zone
for the CuO$_2$ plane.
It looks a truncation of
a
large
Fermi surface
observed in the overdoped cuprates,
and is sometimes called the ``Fermi arc".
More fundamentally,
the normal state of the cuprates remains
a challenge as Mott physics
in the proximity to the Mott insulator\cite{Imada_RMP}.
Although the doped Mott insulators in two dimensions have been 
studied for a long time from various theoretical approaches\cite{Hubbard1,Brinkman,Metzner89,
Meinders93,Lee_RMP,Furukawa,Dzyaloshinskii03,Senechal04,Yang06,Stanescu06,Sakai09,Phillips09},
the nature of the electronic states is not yet fully understood.
Recently revealed pseudogap and arc or pocket of the Fermi surface\cite{Damascelli_RMP,Yoshida06,Meng} require 
a conceptually deeper understandig
for
Mott physics.

In this letter,
we elucidate a key role of
exciton-like physics on this issue.
Excitons are known to be a key concept in physics of
semiconductors\cite{Mahan,Halperin}.
The excitonic state also emerges in the Mott insulator,
for instance as the charge transfer excitation at the optical gap edge
in the cuprates\cite{Uchida91,Ellis,Schuster}, due to a
strong binding
of empty (holon) and doubly occupied (doublon) sites
in half-filled Mott insulators.
In the doped Mott insulators, in spite of screening by doped carriers,
the remnant of the binding may still 
remain as weak binding between a doped holon and the preexisting doublons similarly to excitons.
When an electron or hole is added to the doped Mott insulators, it may appear as 
a normal quasiparticle extended in space.
However, an electron (a hole) can alternatively be added locally to a holon (doublon) site 
with a small cost of the on-site Coulomb repulsion.
This generates a
bound composite particle (cofermion) consisting of
the preexisting holon (doublon) and the added electron (hole).
We call
such cofermions
{\it holo-electron} ({\it doublo-hole}).

A cofermion (a holo-electron or a doublo-hole) dynamically breaks up
into (and recombines from) a conventional quasiparticle and a charge boson.
This dynamical process is naturally interpreted as a 
hybridization between the cofermion
and the quasiparticle.
Here,
we show that the resultant hybridization gap 
offers a natural understanding of a number of key properties 
of the underdoped cuprates\cite{Imada_RMP} such as 
Fermi pocket or
arc formation\cite{Damascelli_RMP, Yoshida06, Meng}, 
psudogap behavior
seen in the single particle spectra\cite{Yoshida06,Ino},
specific heat\cite{Loram01,Momono}, the asymmetric density of states (DOS)\cite{Renner98},
and violation of the Wiedemann-Franz law\cite{Hill01,Proust05},
without any symmetry breaking.
We specifically predict that 
the pseudogap opens as a 
$s$-wave-like
gap in the {\it unoccupied} part above the Fermi level
contrary to the widely assumed $d$-wave structure.

We study the Hubbard
hamiltonian
on a square lattice,
\eqsa{
	\hat{H}=\sum_{i,j}t_{ij}\hatd{c}{i\sigma}\hatn{c}{j\sigma}+U\sum_{i}\hatn{n}{i\uparrow}\hatn{n}{i\downarrow},\label{Hubbard}
}
where $\hatd{c}{i\sigma}$ ($\hatn{c}{i\sigma}$) is
spin-$\sigma$
creation (annihilation) operator
at a site $i$,
while 
$\hatn{n}{i\sigma}$=$\hatd{c}{i\sigma}\hatn{c}{i\sigma}$.
For the hopping $t_{ij}$, we take only $-t$ for the nearest-neighbor and $t'$ for the next-nearest-neighbor pairs.  

We first employ the Kotliar-Ruckenstein
slave-boson formalism\cite{KR}, 
while the local Hilbert space of the Hubbard model is expanded 
not by the original electron $\hat{c}_{\sigma}$ but instead by
introducing
a fermion $\hatn{f}{\sigma}$, which stands for the $\sigma$-spin
quasiparticle,
following
Ref.\onlinecite{Lechermann} and
one slave boson for each Fock state as
$\hat{e}$ for
the empty state (holon) $|0\rangle$, $\hat{p}_{\sigma}$ for
the singly occupied state $|$$\sigma\rangle$ ($\sigma$=$\uparrow$, or $\downarrow$),
and  $\hat{d}$ for the doubly occupied state (doublon) $|$$\uparrow\downarrow\rangle$.
After the mapping,
the
Coulomb repulsion
$U$ is now interpreted as
a ``chemical potential" for
$\hatd{d}{i}$, 
while the correlation
now 
appears, as we describe below, in
hopping process of
$\hatd{f}{i\sigma}$
disturbed by slave-boson motion
under the
local constraints
$\hatd{e}{i}\hatn{e}{i}$+$\sum_{\sigma}\hatd{p}{i\sigma}\hatn{p}{i\sigma}$+$\hatd{d}{i}\hatn{d}{i}$=$1$
and
$\hatd{f}{i\sigma}\hatn{f}{i\sigma}$=$
\hatd{p}{i\sigma}\hatn{p}{i\sigma}$+$\hatd{d}{i}\hatn{d}{i}$
imposed to keep
consistency between the
boson and fermion Hilbert space.
In the enlarged Hilbert space,
these two constraints are assured respectively by the
Lagrange multipliers $\lambda_{i}^{(1)}$ and $\lambda_{i\sigma}^{(2)}$
in the Lagrangian as
\eqsa{
	\hat{\mathcal{L}}=
	\sum_{ij}
	\hatd{f}{i\sigma}(\tau)
	[
	\hat{D}_{i}\delta_{ij}
	+
	\hatn{\zeta}{ij\sigma}(\tau)t_{ij}
	]
	\hatn{f}{j\sigma}(\tau)
	+
	\hat{\mathcal{L}}_{{\rm B}}^{(0)},
	\label{KR_Hubbard}
}
where $\hat{D}_{i}$=$\partial_{\tau}-\mu$+$\lambda_{i\sigma}^{(2)}$,
$\hatn{\zeta}{ij\sigma}(\tau)$=$\hatn{z}{i\sigma}(\tau)\hatd{z}{j\sigma}(\tau)$,
and
$\hatn{z}{i\sigma}
	$=$
	{\hat{g}_{i\sigma}}^{(1)}
	(\hatd{p}{i\sigma}\hatn{e}{i}+\hatd{d}{i}\hatn{p}{i\overline{\sigma}}
	)
	{\hat{g}_{i\sigma}}^{(2)}$ (
$	
{\hat{g}_{i\sigma}}^{(1)}$=$
	(1-
	\hatd{p}{i\overline{\sigma}}\hatn{p}{i\overline{\sigma}}-\hatd{e}{i}\hatn{e}{i}
	)^{-1/2}
$
and 
$	
	{\hat{g}_{i\sigma}}^{(2)}$=$
	(
	1-\hatd{p}{i\sigma}\hatn{p}{i\sigma}-\hatd{d}{i}\hatn{d}{i}
	)^{-1/2}
,$
following the literature\cite{KR}).
A part of the Lagrangian
$\hat{\mathcal{L}}_{{\rm B}}^{(0)}$ contains
$\lambda_{i}^{(1)}$, $\lambda_{i\sigma}^{(2)}$
and
quadratic terms of bosonic fields only\cite{KR} as,
$
	\hat{\mathcal{L}}_{{\rm B}}^{(0)}
	$=$
	\sum_{i}
	\Large{\{}
	\wdtd{e}{i}(\tau)[\partial_{\tau}+\lambda_{i}^{(1)}]\wdtn{e}{i}(\tau)
	+
	\sum_{\sigma}\wdtd{p}{i\sigma}(\tau)[\partial_{\tau}+\lambda_{i}^{(1)}-\lambda_{i\sigma}^{(2)}]
	\wdtn{p}{i\sigma}(\tau)
	+
	\wdtd{d}{i}(\tau)[\partial_{\tau}+U+\lambda_{i}^{(1)}-\sum_{\sigma}\lambda_{i\sigma}^{(2)}
	]\wdtn{d}{i}(\tau)
	\}.
$
To take into account the Gaussian fluctuations of 
the bosonic fields beyond the mean-field level\cite{Castellani92},
the Bogoliubov prescription
is useful, where
the boson operators $\hatn{b}{i}$ ($b=e,d$ or $p_{\sigma}$) are divided into condensate $\cond{b}$
and fluctuating components $\wdtn{b}{i}$
as 
$
	\hatd{b}{i}$=$\cond{b}$+$\wdtd{b}{i}$, $\hatn{b}{i}$=$\cond{b}$+$\wdtn{b}{i}
$.

In this letter, we further make a progress by considering 
low-energy dynamics arising from 
coupled bosons and fermions.
%
First, we reexamine
the strong couling
($U/t\rightarrow +\infty$) limit\cite{Meinders93,Phillips09},
where
adding a $\sigma$-spin electron is possible only to
{$|$$0\rangle$,}
namely, a holon site,
to avoid creating
{$|$$\uparrow\downarrow\rangle$}
(doublon) with
the cost of
$U$.
Creation
operators for the
electron at
{$|$$0\rangle$}
are
given by composite fermion operators
$\hatn{e}{i}\hatd{f}{i\sigma}$.

When $t/U$ becomes nonzero, an added electron may become
a coherent {quasiparticle.}
However, we still have rather localized character of holons,
and it allows alternatively
forming
a collective excitation of a hole and the added electron similarly to an exciton.
{T}his collective character is clearly distinguished from
the conventional {quasiparticle.}
In fact this composite fermion does not have charge in contrast to the 
{quasiparticle.}
Our crucial step is to include dynamics of this composite fermion
expressed by 
$\wdtn{e}{i}\hatd{f}{i\sigma}$ ($\hatd{f}{i\sigma}\wdtn{e}{i}$). 

When we impose
the local constraints
more strictly
for fluctuating bosons beyond the mean-field level, 
it turns out
that,  in the hopping process of $\hat{f}_{\sigma}$ expressed by
$\hat{f}^{\dagger}_{i\sigma}\hat{\zeta}_{ij\sigma}t_{ij}\hat{f}_{j\sigma}$,
the
coefficient {$
	\hat{\zeta}_{ij\sigma}^{(1)}$=$
	g_{1\sigma}^{2}
	g_{2\sigma}^{2}
	(\wdtd{p}{i\sigma}\wdtn{e}{i}+\wdtd{d}{i}\wdtn{p}{i\overline{\sigma}})
	(\wdtd{e}{j}\wdtn{p}{j\sigma}+\wdtd{p}{j\overline{\sigma}}\wdtn{d}{j})
$}
is dominating (see Appendix. \ref{A1b}).
Here we employ
$g_{1\sigma}$=$(1-\overline{p}_{0\overline{\sigma}}^{2}
-\overline{e}^{2}_0)^{-1/2}$ and $g_{1\sigma}$=$(1-\overline{p}_{0\sigma}^{2}
-\overline{d}^{2}_0)^{-1/2}$ following Ref.\onlinecite{KR}.
This vertex stands for
the {\it backflow} consisting of bosons,
originating from the
{quasiparticle}
motions.

Then we treat
coupling of charge bosons and
{quasiparticles}
in $\hatd{f}{i\sigma}\hat{\zeta}_{ij\sigma}^{(1)}\hatn{f}{j\sigma}$,
which represents a part of the electron-electron interactions
in the hamiltonian (\ref{Hubbard}),
by interpreting the form such as
$\hatd{f}{i\sigma}
   \wdtn{b}{i}\wdtd{b}{j}
\hatn{f}{j\sigma}$ as the decoupled product of 
$\hat{\mbox{\boldmath$C$}}_{i\sigma}^{\dagger}
$=$
(\wdtn{e}{i},\wdtd{d}{i}
)\hatd{f}{i\sigma}$ 
and 
$\hat{\mbox{\boldmath$C$}}_{i\sigma}^{\ }
$=$\hatn{f}{i\sigma}
(\wdtd{e}{i},\wdtn{d}{i}
)^{T} $.
Namely, this boson-{quasiparticle}
interaction is equivalently
treated by 
introducing integrals over the  grassmanian Stratonovich-Hubbard fields 
$\hat{\mbox{\boldmath$\Upsilon $}}_{i\sigma}^{\dagger}
$=$
(\hatd{\psi}{i\sigma}, \hatd{\chi}{i\sigma}
)$
and
$\hat{\mbox{\boldmath$\Upsilon $}}_{i\sigma}^{\ }
$=$
(\hatn{\psi}{i\sigma}, \hatn{\chi}{i\sigma}
)^{T}$
where the interaction is formally transformed to
the hybridization of $\hat{\mbox{\boldmath$\Upsilon $}}$ and $\hat{\mbox{\boldmath$C $}}$
as 
$\hat{\mbox{\boldmath$C$}}_{i\sigma}^{\dagger}
\hat{\mbox{\boldmath$C$}}_{j\sigma}^{\ } 
\rightarrow
\hat{\mbox{\boldmath$C$}}_{i\sigma}^{\dagger}
\hat{\mbox{\boldmath$\Upsilon $}}_{j\sigma}
+
\hat{\mbox{\boldmath$\Upsilon $}}_{i\sigma}^{\dagger}
\hat{\mbox{\boldmath$C$}}_{j\sigma}
-
\hat{\mbox{\boldmath$\Upsilon $}}_{i\sigma}^{\dagger}
\hat{\mbox{\boldmath$\Upsilon $}}_{j\sigma}
$ (see Appendix. \ref{A1c}).

The newly introduced Grassmann fields $\hatd{\psi}{i\sigma}$ and $\hatn{\chi}{i\sigma}$ are
physically interpreted as cofermions, the holo-electron and doublo-hole, respectively.
Then,
$\hat{\mbox{\boldmath$C$}}_{i\sigma}^{\dagger}
\hat{\mbox{\boldmath$\Upsilon $}}_{j\sigma}
$+$\hat{\mbox{\boldmath$\Upsilon $}}_{i\sigma}^{\dagger}\hat{\mbox{\boldmath$C$}}_{j\sigma}$
is naturally interpreted as
breakup and recombination processes of the cofermions.
After integrating fluctuating bosons out, as is mentioned below, it results
in the hybridization between cofermions $\hat{\psi}_{\sigma}$, $\hat{\chi}_{\sigma}$ and quasiparticles $\hat{f}_{\sigma}$.

\begin{figure}[t]
\begin{center}
\includegraphics[width=8.5cm]{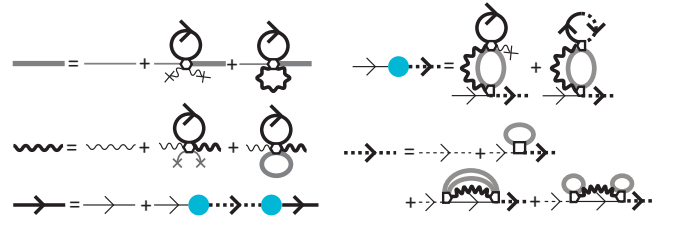}
\end{center}
\caption{
Diagrams for the Dyson equations.
Solid and dashed lines with arrows represent propagators of the
quasiparticles
and cofermions, respectively.
Wavy lines stand for the charge bosons,
and
solid lines
are the spin bosons.
Condensations of bosons
are represented by lines terminated at crosses.
Coupling constant $g_{1\sigma}^{2}g_{2\sigma}^{2}t_{ij}$
is represented by open polygons.
Here,
we do not distinguish
holons and doublons.
Spins and ``flavors" of cofermions $(\psi,\chi)$ are also not distinguished in the diagram,
for simplicity.
Filled circles stand for the amplitude of the hybridization between quasiparticles and cofermions.
\label{Dyson}}
\end{figure}
We treat Gaussian fluctuations of
bosons,
and the dynamical coupling between quasiparticles and cofermions by using a set of the Dyson equations
up to the second order of $t_{ij}$,
as is depicted in Fig.\ref{Dyson}:
Thick lines (thick wavy lines)
stand for the dressed Green's functions of the charge bosons
$
\mathcal{A}^{ab}(r,\tau)$=$-\langle T\beta_{i}^{a}(\tau){\beta_{j}^{b}}^{\dagger}(0)\rangle
$
(spin bosons
$
\mathcal{C}^{ab}(r,\tau)$=$-\langle T\phi_{i}^{a}(\tau){\phi_{j}^{b}}^{\dagger}(0)\rangle
$),
where $a,b$=$1,2$, $r$=$i$$-$$j$, $(\beta_{i}^{1},\beta_{i}^{2})$=$(\wdtn{e}{i},\wdtn{d}{i})$,
and $(\phi_{i}^{1},\phi_{i}^{2})$=$(\wdtn{p}{i\sigma},\wdtd{p}{i\overline{\sigma}})$.
Here we neglect the coupling between charge and spin bosons
such as $\langle\wdtd{p}{i\sigma}\wdtn{e}{i}\rangle$, which
vanishes
in Mott insulators
and gives higher order terms scaled by the doping rate $x$ for $|x|$$\ll$1.
Thick lines with arrows represent
the dressed quasiparticles $\mathcal{G}_{\sigma}^{(f)}(r,\tau)$.
Thin lines (thin wavy lines) represent
bare propagators of the charge bosons $\mathcal{A}_{0}^{ab}(r,\tau)$ (spin bosons $\mathcal{C}_{0}^{ab}(r,\tau)$), determined by $\hat{\mathcal{L}}_{{\rm B}}^{(0)}$.
Thin lines with arrows are bare propagators of the quasiparticles
$\mathcal{G}_{0\sigma}^{(f)}(r,\tau)$ determined by
$
	\hat{\mathcal{L}}_{0}$=$\sum_{ij}
	\hatd{f}{i\sigma}(\tau)
	[
	\hat{D}_{i}\delta_{ij}
	+
	\zeta_{0\sigma}t_{ij}
	]
	\hatn{f}{j\sigma}(\tau),
$
where $\zeta_{0\sigma}$=$g_{1\sigma}^{2}g_{2\sigma}^{2}
(\conp{p}{\sigma}\cond{e}+\cond{d}\conp{p}{\overline{\sigma}})^{2}$.
The Lagrangian $\hat{\mathcal{L}}_{0}$ is obtained by decoupling the fluctuating bosons
from the Lagrangian $\hat{\mathcal{L}}-\hat{\mathcal{L}}^{(0)}_{{\rm B}}$ (see Eq.(\ref{KR_Hubbard})).
Thick and thin dashed lines with arrows are the cofermion propagators $\mathcal{F}^{cd}$ $(c,d$=$\psi,\chi)$ and {$\mathcal{F}^{cd}_{0}$=$\delta_{cd}/\epsilon$ $(\epsilon\rightarrow 0)$}, respectively.
This peculiar divergence of $\mathcal{F}^{cd}_{0}$ is because
the Lagrangian does not include cofermions if we neglect the interactions between
bosons and fermions.
By solving the Dyson equations,
we obtain the propagators for the quasiparticles and cofermions.
Here the bosonic degrees of freedom are taken into account in a self-consistent fashion, 
through the cofermion self-energy
and the amplitude
of hybridization
between the quasiparticles and cofermions.

Now we show how our
self-consistent solution of coupled quasiparticles, bosons and cofermions
predicts normal state properties.
We show the
result
at $U$=$12t$ and $t'$=$0.25t$ 
to get insight into the cuprate superconductors
by restricting to
paramagnetic solutions at temperature $T$=$0$.
First, we give the spectral functions calculated from
the electron Green's function 
$
	G_{\sigma}(k,\omega)
$
(see Appendix. \ref{A1d});
$
	A(k,\omega)$=$-{\rm Im}\left[G_{\sigma}(k,\omega)\right]/\pi.
$
\begin{figure}[t]
\begin{center}
\includegraphics[width=7.5cm]{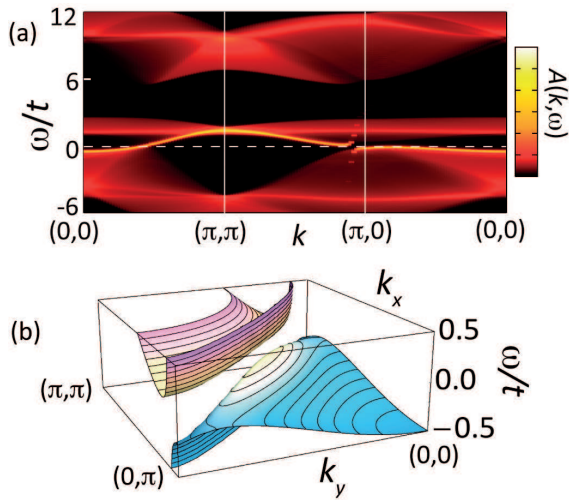}
\end{center}
\caption{
(color online).
Calculated spectral functions for $t'/t$=0.25, $U/t$=12 and $x$=$0.05$.
(a) Spectral function $A(k,\omega)$ along lines running from $(0,0)$,
$(\pi,\pi)$
and $(\pi,0)$ to $(0,0)$.
The dashed line is the Fermi level. We use a finite broadening factor $\delta$=$0.05t$.
(b) Band dispersion of quasiparticle for $x$=$0.05$.
\label{Akw_GL}}
\end{figure}
\begin{figure}[t]
\begin{center}
\includegraphics[width=8.5cm]{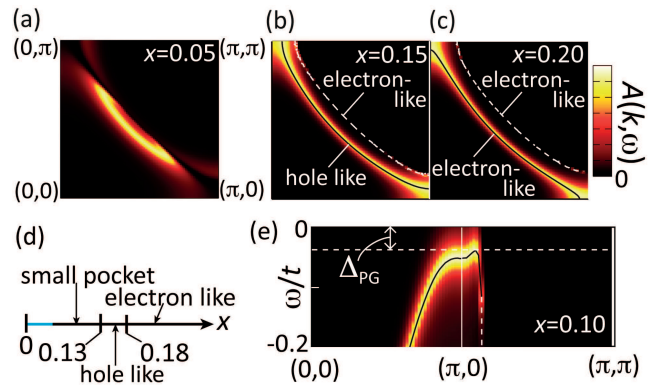}
\end{center}
\caption{
(color online).
Reconstructed Fermi surfaces and spectral functions $A_{f}(k,\omega)$ for $t'/t$=0.25 and $U/t$=12.
(a)-(c) {$A_{f}(k,\omega)(\equiv -{\rm Im}[G_{\sigma}^{(f)}(k,\omega)]/\pi)$}
at $\omega=i\delta$.
Here we take the broadening factor $\delta=0.05t$.
Solid and dashed lines illustrate the poles of {quasiparticles.}
(d) Doping dependence of Fermi-surface topology in our theory.
(e)
{$A_{f}(k,\omega)$} along
the symmetry line running from $(0,0)$ through $(\pi, 0)$ to $(\pi, \pi)$
for $x=0.10$.
Thin solid and dashed white lines illustrate poles of quasiparticles.
We define
$\Delta_{{\rm PG}}$ as the gap between
$\mu$ ($\omega=0$) and the
maximum of the quasiparticle dispersion below $\mu$ along
this symmetry line.
\label{Akw}}
\end{figure}
In Fig.\ref{Akw_GL}a, we show
$A(k,\omega)$
for the hole concentration $x=0.05$.
Two main features are found;
the coherent band arising from the
quasiparticle
around the Fermi level
and
the remnant of the upper (lower) Hubbard band at $\omega$$>$$6t$ $(\omega$$<$$-2t)$
generated by dynamics of $\widetilde{e}$ and $\widetilde{d}$~\cite{Raimondi,Castellani92}.

Here, we focus on reconstructions of the Fermi surface in the coherent band.
The
quasiparticle
Green's function
is given as
\eqsa{
	G_{\sigma}^{(f)}(k,\omega)
	\simeq
	\left[\omega
	-\zeta_{0\sigma}\epsilon_{k}
	+\mu-
	\Sigma_{f}(k,\omega)
	\right]^{-1}
	\label{QDF_F1},
}
where $\Sigma_{f}$=$\Delta(k)^{2}/(
	\gamma_{k}
	\omega
	-\alpha_{k})$
is the quasiparticle self-energy arising from the quasiparticle-cofermion hybridization $\Delta(k)$.
Here the cofermion propagator
$(\gamma_{k}
	\omega
	-\alpha_{k})^{-1}$ is obtained from the expansion around the cofermion pole
(see Appendix. \ref{A1d}).
The quasiparticles Green's function
has a hybridization gap
due to the hybridization with the cofermions and
$G_{\sigma}^{(f)}$
shows the divergence of
the
{quasiparticle} self-energy
given by $\Sigma_{f}$
at the zero surface\cite{Dzyaloshinskii03,Yang06,Stanescu06,Sakai09}
defined by
$\gamma_{k}\omega-\alpha_{k}$=$0$.
In our theory,
the zero surface splits the band dispersion
and generates a distinct $s$-wave-like
gap (as is seen in Fig.\ref{Akw_GL}b
and
supported by recent numerical observation\cite{Stanescu06,Sakai09}).
For small hole doping such as $x$=$0.05$,
our theory predicts that the reconstructed Fermi surface becomes
a small pocket
as we see in Fig.\ref{Akw}a-\ref{Akw}c.

Actually, the topological transitions occur at $x$$\simeq$$0.13$ and $x$$\simeq$$0.18$
(see Fig.\ref{Akw}d).
Below $x$$\simeq$$0.13$,
the Fermi surface consists of small hole pockets.
It is difficult to distinguish the pockets from the arc structure as we see in Fig.\ref{Akw}a.
This is because 
the zero surface near the outer part partially destroys the quasiparticles.
For $0.13$$\lesssim$$x$,
large Fermi surfaces appear, instead of Fermi pockets.
For $0.13$$\lesssim$$x$$\lesssim$$0.18$,
a hole-like surface centered at $(\pi,\pi)$,
and
an electron-like one centered at $(\pi,\pi)$
coexist
(see Fig.\ref{Akw}b).
However, the electron-like surface is hardly seen again
because of the nearby zero surface.
Our
result
shows a nontrivial topological transition at $x$$\simeq$$0.13$ as
a consequence of the hybridization with cofermions.

A gap $\Delta_{{\rm PG}}$ measured from the Fermi level {$\mu$}
emerges near the antinode,
corresponding to
the pseudogap in the ARPES
as we identify in Fig.\ref{Akw}e.
\begin{figure}[t]
\begin{center}
\includegraphics[width=8.5cm]{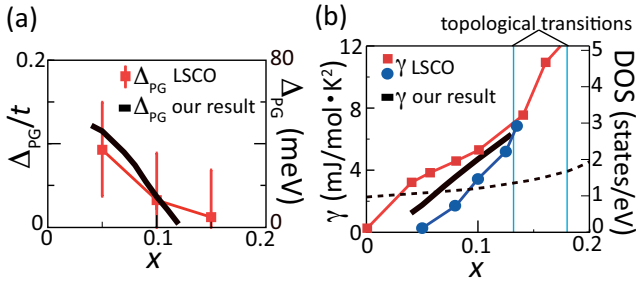}
\end{center}
\caption{
(color online).
(a)
Pseudogap $\Delta_{{\rm PG}}$ as a function of hole doping rate $x$. Thick solid line is
$\Delta_{{\rm PG}}$
calculated by our theory.
Closed (red) squares show
$\Delta_{{\rm PG}}$ estimated from the ARPES
\cite{Ino}
for La$_{2-x}$Sr$_x$CuO$_4$ (LSCO).
(b)
DOS of electrons 
vs. $x$ in the present theory (bold solid curve).
The thin dashed curve stands for the DOS of the non-interacting case.
Closed (blue) circles are obtained by a linear extrapolation 
of low-temperature normal-state $\gamma$ of LSCO in Ref.\onlinecite{Loram01} to
$T$$\rightarrow$$0$,
representing the presumable lower limit.
Closed (red) squares stand for 
the estimate for LSCO obtained in Ref.\onlinecite{Momono} by the Zn doping,
which may be an upper limit.
All of the present results are obtained at $t$=0.4eV,
$t'/t$=0.25 and $U/t$=12.
\label{PG_DOS}}
\end{figure}
The pseudogap $\Delta_{{\rm PG}}$ is determined by
the hybridization gap $\Delta (k)$, basically scaled by
a fraction of $t$,
consistently
with numerical studies\cite{Senechal04}{.}
The doping dependence of
$\Delta_{{\rm PG}}$
is given in Fig.\ref{PG_DOS}a, in agreement with 
the ARPES
for LSCO\cite{Ino,Yoshida06}.

For $x$$\lesssim$$0.13$,
the density of states (DOS) of the electrons $\hatn{c}{k\sigma}$ at the Fermi level,
$\rho_{{\rm F}}$$\simeq$$-\zeta_{0}\int d^{2}k A_{f}(k,0)/4\pi^3$,
is clearly
suppressed, as is
illustrated in Fig.\ref{PG_DOS}b.
We compare $\rho_{{\rm F}}$ with {the} specific heat coefficient $\gamma$
measured for
LSCO\cite{Momono,Loram01}
by using the conventional relation
$\gamma$=$\pi^{2}\rho_{F}/3
$ at $T$=$0$.
Our
$\gamma$ is consistent with experiments.
The $\omega$-dependence of the DOS 
shows significant asymmetry around $\omega$$=$$0$
in contrast
to the DOS for the non-interacting case (see Appendix. \ref{A2}).
This asymmetry of the DOS naturally explains
the asymmetric
tunneling spectra
observed
in the hole-underdoped cuptrates\cite{Renner98}
(see Refs.\onlinecite{Anderson67} and \onlinecite{Nieminen09} for
different interpretations).

The present result slightly depends on the choice of the parameters.
For instance, $\Delta_{{\rm PG}}$ decreases from the present result 
by an amount $\sim$$0.05t$ at $t'/t$=0.25 and $U/t$=15
or $t'/t$=0.15 and $U/t$=12, while the qualitative features are robust.

Here we note the difference of the cofermion from the spinon 
\cite{Lee_RMP}:
Although the cofermion carries a spin but no charge as in the spinon, 
cofermions coexist with quasiparticles in different part of energy-momentum
space as the electron differentiation in contrast to the spinons.

The present cofermion contributes to the entropy and
the thermal conductivity $\kappa$ in addition to the quasiparticle.
On the other hand, 
the electric conductivity $\sigma$ is contributed only from the quasiparticle.
Therefore
we expect a serious breakdown of the 
Wiedeman-Franz
law\cite{Hill01,Proust05} that
predicts a universal constant $L_{0}$=$\pi^{2}k_{{\rm B}}^{2}/3e^{2}$
for the ratio $L$$\equiv$$\kappa/T\sigma$.
Our theory
predicts $L>L_{0}$.

We propose
to test
our specific prediction of the
$s$-wave-like
gap structure
in unoccupied spectra, for example,
by improving the low-energy electron spectroscopies,
such as the inverse photoemission,
the low-energy electron diffraction spectroscopy{,}
resonant inelastic X-ray spectroscopy{,}
or time resolved photoemission spectroscopy.
The mid-infrared peak and
long tail of
{the optical conductivity\cite{Uchida91}}
indeed supports our prediction.

Our finding is
that hidden cofermionic particles
called
{\it holo-electrons} and {\it doublo-holes}
play a key role:
The cofermions hybridize with the
quasiparticles
and cause a hybridization gap 
identified as the pseudogap.
A number of resultant properties consistent
with the unusual normal states of
the cuprates
support
relevance of our cofermion theory to
physics of the cuprates.

The authors thank Yukitoshi Motome and Shiro Sakai for useful discussions.
Y.Y. is supported by the Japan Society for the Promotion of Science.

\appendix*
\section{}
In this Appendix, we show
theoretical details
on introducing the cofermions, construction of quasiparticle Green's functions,
and supplementary results on the quasiparticle density of states.
\subsection{Details of theory}
	As our {theoretical} starting point,
we employ the Kotliar-Ruckenstein
slave-boson formalism\cite{KR} for the Hubbard model, 
where the local Hilbert space of the Hubbard model is expanded by
introducing one slave boson for each Fock state as
$\hat{e}$ for
the empty state (holon) $|0\rangle$, $\hat{p}_{\sigma}$ for
the singly occupied state $|$$\sigma\rangle$ ($\sigma$=$\uparrow$,or $\downarrow$),
and  $\hat{d}$ for the doubly occupied state (doublon) $|$$\uparrow\downarrow\rangle$.
In addition to these bosons {$\hat{b}$ ($b=e, p_{\sigma}$ or $d$)},
fermion operator $\hatn{f}{\sigma}$ is introduced to stand for the $\sigma$-spin
{QP}. 
The mapping between the original electrons and $\hat{f}$
combined with {$\hat{b}$} is given by 
$
	\hatd{c}{i\sigma}\doteq
	\hatn{z}{i\sigma}
	\hatd{f}{i\sigma},
$
where $\hatn{z}{i\sigma}$ is defined\cite{KR,Raimondi} as
\eqsa{
	\hatn{z}{i\sigma}
	=
	{\hat{g}_{i\sigma}}^{(1)}
	(\hatd{p}{i\sigma}\hatn{e}{i}+\hatd{d}{i}\hatn{p}{i\overline{\sigma}}
	)
	{\hat{g}_{i\sigma}}^{(2)}
}
with
\eqsa{
	{\hat{g}_{i\sigma}}^{(1)}=
	(1-
	\hatd{p}{i\overline{\sigma}}\hatn{p}{i\overline{\sigma}}-\hatd{e}{i}\hatn{e}{i}
	)^{-1/2},
}
and 
\eqsa{
	{\hat{g}_{i\sigma}}^{(2)}=
	(
	1-\hatd{p}{i\sigma}\hatn{p}{i\sigma}-\hatd{d}{i}\hatn{d}{i}
	)^{-1/2}
,}
following the literature\cite{KR}.

We need to impose local
constraints to eliminate unphysical states.
First, only one boson should occupy each local state as
\eqsa{
	\hatd{e}{i}\hatn{e}{i}+\sum_{\sigma}\hatd{p}{i\sigma}\hatn{p}{i\sigma}+\hatd{d}{i}\hatn{d}{i}=1.
}
Second, the number operator of $\hat{f}_{\sigma}$
is necessarily given as  
\eqsa{
	\hatd{f}{i\sigma}	\hatn{f}{i\sigma}=\hatd{p}{i\sigma}\hatn{p}{i\sigma}+\hatd{d}{i}\hatn{d}{i}.
}
These constraints 
are represented by integrals over
Lagrange multipliers $\lambda_{i}^{(1)}$,
and $\lambda_{i\sigma}^{(2)}$,
in the path integral of the
Lagrangian discussed below. 

In the expanded Hilbert space,
the Lagrangian of
the Hubbard Hamiltonian is
mapped to
\eqsa{
	\hat{\mathcal{L}}=
	\sum_{ij}
	\hatd{f}{i\sigma}(\tau)
	[
	\hat{D}_{i}\delta_{ij}
	+
	\hatn{\zeta}{ij\sigma}(\tau)t_{ij}
	]
	\hatn{f}{j\sigma}(\tau)
	+
	\hat{\mathcal{L}}_{{\rm B}}^{(0)},
	\label{KR_Hubbard_2}
}
where $\hat{D}_{i}=\partial_{\tau}-\mu+\lambda_{i\sigma}^{(2)}$
and
$\hatn{\zeta}{ij\sigma}(\tau)=\hatn{z}{i\sigma}(\tau)\hatd{z}{j\sigma}(\tau)$.
$\hat{\mathcal{L}}_{{\rm B}}^{(0)}$ contains
$\lambda_{i}^{(1)}$, $\lambda_{i\sigma}^{(2)}$
and
quadratic terms of bosonic fields only\cite{KR} as,
\eqsa{
	\hat{\mathcal{L}}_{{\rm B}}^{(0)}
	&=&
	\sum_{i}
	\Large{\{}
	\wdtd{e}{i}(\tau)[\partial_{\tau}+\lambda_{i}^{(1)}]\wdtn{e}{i}(\tau)
	\nn
	&+&
	\sum_{\sigma}\wdtd{p}{i\sigma}(\tau)[\partial_{\tau}+\lambda_{i}^{(1)}-\lambda_{i\sigma}^{(2)}]
	\wdtn{p}{i\sigma}(\tau)
	\nn
	&+&
	\wdtd{d}{i}(\tau)[\partial_{\tau}+U+\lambda_{i}^{(1)}-\sum_{\sigma}\lambda_{i\sigma}^{(2)}
	]\wdtn{d}{i}(\tau)
	\}.
}
The
on-site
Coulomb 
$U$ is {now interpreted as}
a ``chemical potential" for
$\hatd{d}{i}$, 
while the correlation
now 
{appears in}
hopping process of
$\hatd{f}{i\sigma}$
disturbed by slave boson motion.
	\subsubsection{Bosonic fluctuations}\label{A1b}
	To take into account the Gaussian fluctuations of 
the bosonic fields beyond the mean-field level\cite{Castellani92},
the Bogoliubov prescription
is useful, where
the boson operators are divided into condensate components $\cond{b}$
and fluctuating components $\wdtn{b}{i}$
as 
$
	\hatd{b}{i}=\cond{b}+\wdtd{b}{i},\ \hatn{b}{i}=\cond{b}+\wdtn{b}{i}
$
with $b=e,d$ or $p_{\sigma}$.

\begin{figure}[h]
\begin{center}
\includegraphics[width=6cm]{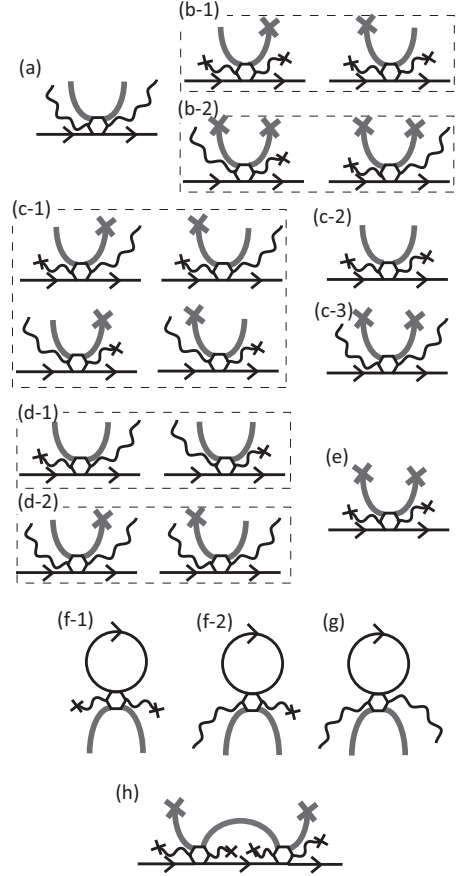}
\end{center}
\caption{(a)-(e) Diagrams representing terms in $\hatd{f}{i\sigma}\hat{\zeta}_{ij}\hatn{f}{j\sigma}$.
(f)-(h), Examples of the diagrams
for perturbation
generated from
the same term as (a)-(e).
Solid lines with arrows represent propagators of the
quasiparticles.
Wavy lines stand for the charge bosons,
and solid lines
are the spin bosons.
Condensations of bosons
are represented by lines terminated at crosses.
Coupling constant $g_{1\sigma}^{2}g_{2\sigma}^{2}t_{ij}$
is represented by open polygons.
Here,
we do not distinguish
holons and doublons.
Spins are also not distinguished in the diagram,
for simplicity.
\label{diagrams_gs}}
\end{figure}

When we impose
the local constraints
more strictly
for fluctuating bosons beyond the mean-field level, 
it turns out that the term
\eqsa{
	\hat{\zeta}_{ij\sigma}^{(1)}=
	g_{1\sigma}^{2}
	g_{2\sigma}^{2}
	(\wdtd{p}{i\sigma}\wdtn{e}{i}+\wdtd{d}{i}\wdtn{p}{i\overline{\sigma}})
	(\wdtd{e}{j}\wdtn{p}{j\sigma}+\wdtd{p}{j\overline{\sigma}}\wdtn{d}{j})
}
represented by the diagram in Fig.\ref{diagrams_gs}a
is dominating among all the possible diagrams for $\hat{\zeta}_{ij\sigma}$ in Eq.(\ref{KR_Hubbard_2}).
{Here we employ $g_{1\sigma}^{2}=(1-\overline{p}_{0\overline{\sigma}}^{2}
-\overline{e}^{2}_0)^{-1}$ and $g_{1\sigma}^{2}=(1-\overline{p}_{0\sigma}^{2}
-\overline{d}^{2}_0)^{-1}$ by following Ref.\onlinecite{KR}.}

To elucidate why we retain $\hat{\zeta}_{ij\sigma}^{(1)}$,
we
classify
the diagrams illustrated in Fig.\ref{diagrams_gs}
into four types,
categorized by
time dependence (or frequency dependence)
of quasiparticles and the local conservation
of the boson densities:
\begin{description}
\item[\textsf{(T-1)}]
Diagrams
containing external propagators of quasiparticles,
in addition to
bosonic propagators
violating the local conservation
of boson densities
(Figures \ref{diagrams_gs}b,\ref{diagrams_gs}c,\ref{diagrams_gs}d, and \ref{diagrams_gs}h).
Here the violation means
that before and after
the interactions (represented by hexagons),
the number of bosons expressed
by external boson propagators
is not the same.
\item[\textsf{(T-2)}]
Diagrams that contain
time dependence
of quasiparticles,
but that do not violate the local conservation
(Figures \ref{diagrams_gs}a and \ref{diagrams_gs}e).
\item[\textsf{(T-3)}]
Diagrams that do not contain
time dependence
of quasiparticles
but do
violate the local conservation
(Fig.\ref{diagrams_gs}f).
\item[\textsf{(T-4)}]
Diagrams that
neither
include
time dependence
of quasiparticles
nor
violate the local conservation
(Fig.\ref{diagrams_gs}g).
\end{description}

{Here} we present our guiding principle to
take account of
boson fluctuations:
We exclude ({\bf \textsf{T-1}}) because it violates
the local conservation when the quasiparticles
dynamically fluctuate.
On the other hand we retain diagrams
belonging to the categories
({\bf \textsf{T-2}}), ({\bf \textsf{T-3}}), and ({\bf \textsf{T-4}}).
The reason to retain these diagram is as follows.
The diagrams in the category
({\bf \textsf{T-2}}) do not violate the local conservation,
when bosons fluctuate.
Therefore, we take the diagrams in this category
into account.
On the other hand,
the diagrams in the category
({\bf \textsf{T-3}}) do violate the local conservation.
However,
in these diagrams, quasiparticles
enter as time averaged Green's functions.
Therefore,
quasiparticles feel the time averaged bosonic motions.
The real violation of the local conservation
occurs only when a dynamical quasiparticle process
is induced by
fluctuating boson hoppings.
On the contrary,
the real
violation
does not occur when the quasiparticles emerge as
the time averaged quantities
as in the case of ({\bf \textsf{T-3}}).
This is the reason to retain the diagrams in the category ({\bf \textsf{T-3}}).
Since ({\bf \textsf{T-4}}) does not violate local conservation, we retain it.

For the slave-particle formalism of correlated fermion systems,
it is well-known that
fluctuations of
gauge fields play an important role on
reinforcing the local constraint imposed on slave particles\cite{Lee_RMP}.
It was pointed out by Jolicoeur and Le Guillou that
the Kotliar-Ruckenstein formalism has
the $U$(1)$\times$$U$(1)$\times$$U$(1)
gauge symmetry\cite{gauge_KR}.
It comes from the phase symmetry of the slave bosonic particles, namely, $\hatn{e}{i}$, $\hatn{p}{i\sigma}$, and
$\hatn{d}{i}$.

In our theory, we will treat fluctuations of such phases together with fluctuations of the amplitude of the
condensation fraction of these slave particles, by using the Bogoliubov prescription.
Therefore, the phase fluctuations are taken into account, although 
the $U$(1)$\times$$U$(1)$\times$$U$(1)
gauge structure is not strictly conserved.
	\subsubsection{{Stratonovich-Hubbard trasformation}}\label{A1c}
	We introduce Grassmannian valuables (or fermionic fields)
$\hat{\mbox{\boldmath$\Upsilon$}}_{i\sigma}^{\ }
	=
	(\hatn{\psi}{i\sigma}, \hatn{\chi}{i\sigma})^{T}$
that stands for the cofermions as are discussed in the main article,
by using
a following identity
\eqsa{
	\int
	\prod_{i\sigma}
	d\hat{\mbox{\boldmath$\Upsilon$}}_{i\sigma}^{\dagger}
	d\hat{\mbox{\boldmath$\Upsilon$}}_{i\sigma}^{\ }
	e^{
	\mathcal{A}
	}
	=
	\det
	\left[
	\widetilde{\mbox{\boldmath$T$}}_{\uparrow}\widetilde{\mbox{\boldmath$T$}}_{\downarrow}
	\right],\label{Id}
}
where matrices $\widetilde{\mbox{\boldmath$T$}}_{\sigma}$ are defined as
\eqsa{
(\widetilde{\mbox{\boldmath$T$}}_{\sigma})_{ij}={g_{1\sigma}^{2}g_{2\sigma}^{2}}{t_{ij}}
\left[
\begin{array}{cc}
\wdtd{p}{i\sigma}\wdtn{p}{j\sigma}&\wdtd{p}{i\sigma}\wdtd{p}{j\overline{\sigma}}\\
\wdtn{p}{i\overline{\sigma}}\wdtn{p}{j\sigma}&\wdtn{p}{i\overline{\sigma}}\wdtd{p}{j\overline{\sigma}}
\end{array}
\right],
}
and
\eqsa{
	\mathcal{A}
	&=&
	\int_{0}^{\beta}d\tau
	\sum_{ij\sigma}
	\left[
	\left(\hat{\mbox{\boldmath$\Upsilon$}}_{i\sigma}^{\dagger}(\tau)
	-
	\hat{\mbox{\boldmath$C$}}_{i\sigma}^{\dagger}(\tau)
	\right)
	\right.
	\nn
	&\times&
	\left.
	(\widetilde{\mbox{\boldmath$T$}}_{\sigma})_{ij}
	\left(\hat{\mbox{\boldmath$\Upsilon$}}_{j\sigma}^{\ }(\tau)
	-
	\hat{\mbox{\boldmath$C$}}_{j\sigma}^{\ }(\tau)
	\right)
	\right].
}
Here we use vector notations as are defined in the main article as
\eqsa{
	\hat{\mbox{\boldmath$C$}}_{i\sigma}^{\dagger}
	&=
	\left(\wdtn{e}{i},\wdtd{d}{i}\right)\hatd{f}{i\sigma},\ 
	\hat{\mbox{\boldmath$C$}}_{i\sigma}^{\ }
	=
	\hatn{f}{i\sigma}\left(\wdtd{e}{i},\wdtn{d}{i}\right)^{T}.
}

\begin{figure}
\begin{center}
\includegraphics[width=7cm]{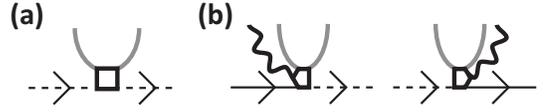}
\end{center}
\caption{
Diagrams for the transformed Lagrangians.
Solid and dashed lines with arrows represent propagators of the
quasiparticles
and cofermions, respectively.
Wavy lines stand for the charge bosons,
and solid lines
are the spin bosons.
(a)
The diagram represents the Lagrangian $\mathcal{L'}_{a1}$ (Eq.(\ref{S_1})).
(b)
The diagrams stand for terms in the Lagrangian $\mathcal{L'}_{a2}$ (Eq.(\ref{S_2})).
\label{diagrams_2}}
\end{figure}

The identity Eq.(\ref{Id}) gives the transformation
for a coupling term of the quasiparticles and fluctuating bosons depicted
in Fig.\ref{diagrams_gs}a,
\eqsa{
	\mathcal{L}_{{\rm a}}
	=
	\sum_{ij\sigma}
	\hat{\mbox{\boldmath$C$}}_{i\sigma}^{\dagger}(\tau)
	(\widetilde{\mbox{\boldmath$T$}}_{\sigma})_{ij}
	\hat{\mbox{\boldmath$C$}}_{j\sigma}^{\ }(\tau),
}
as
\eqsa{
	\exp\left[
	-\int_{0}^{\beta}d\tau
	\mathcal{L}_{{\rm a}}
	\right]
	=
	\frac{
	\displaystyle
	\int
	\prod_{i\sigma}
	d\hat{\mbox{\boldmath$\Upsilon$}}_{i\sigma}^{\dagger}
	d\hat{\mbox{\boldmath$\Upsilon$}}_{i\sigma}^{\ }
	e^{
	-\int_{0}^{\beta}d\tau
	\mathcal{L'}_{a}
	}}{\det [\wdtn{\mbox{\boldmath$T$}}{\uparrow}\wdtn{\mbox{\boldmath$T$}}{\downarrow}]}
,
}
where
\eqsa{
	\mathcal{L'}_{a}
	=
	\mathcal{L'}_{a1}
	+
	\mathcal{L'}_{a2}
	,
}
\eqsa{
	\mathcal{L'}_{a1}
	=
	\sum_{ij\sigma}
	\hat{\mbox{\boldmath$\Upsilon$}}_{i\sigma}^{\dagger}(\tau)
	(\widetilde{\mbox{\boldmath$T$}}_{\sigma})_{ij}
	\hat{\mbox{\boldmath$\Upsilon$}}_{j\sigma}^{\ }(\tau)
	\label{S_1},
}
\eqsa{
	\mathcal{L'}_{a2}
	&=&
	-
	\sum_{ij\sigma}
	\left\{
	\hat{\mbox{\boldmath$C$}}_{i\sigma}^{\dagger}(\tau)
	(\widetilde{\mbox{\boldmath$T$}}_{\sigma})_{ij}
	\hat{\mbox{\boldmath$\Upsilon$}}_{j\sigma}^{\ }(\tau)
	\right.
	\nn
	&+&
	\left.
	\hat{\mbox{\boldmath$\Upsilon$}}_{i\sigma}^{\dagger}(\tau)
	\widetilde{\mbox{\boldmath$T$}}_{ij}
	\hat{\mbox{\boldmath$C$}}_{j\sigma}^{\ }(\tau)
	\right\}	
	.\label{S_2}
}
These transformed Lagrangian $\mathcal{L'}_{a1}$ (Fig.\ref{diagrams_2}a)
and $\mathcal{L'}_{a2}$ (Fig.\ref{diagrams_2}b)
lead to the cofermions' self-energy and the hybridization between the quasiparticles
and cofermions, respectively after integrating out the fluctuating bosonic degrees of freedom.
It will be
discussed below, by using a set of the Dyson equations.
	\subsubsection{{Prescription for self-consistent procedure and Green's functions}}\label{A1d}
\begin{figure}[h]
\begin{center}
\includegraphics[width=6cm]{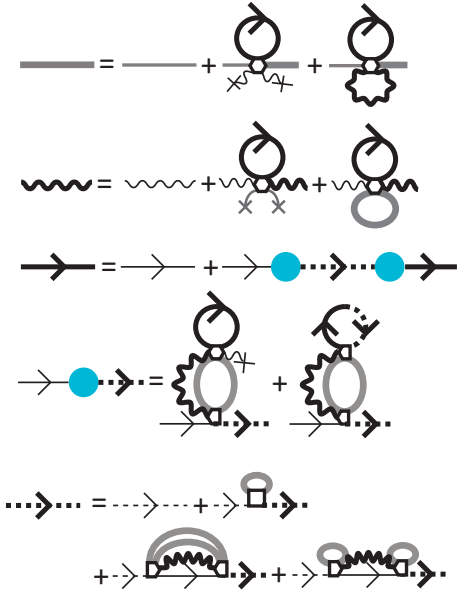}
\end{center}
\caption{
(Same figure as Fig.1 in the main article)
Diagrams for the Dyson equations.
Solid and dashed lines with arrows represent propagators of the
quasiparticles
and cofermions, respectively.
Thin wavy lines stand for the charge bosons,
and thick solid lines
are the spin bosons.
Condensations of bosons
are represented by lines terminated at crosses.
Coupling constant $g_{1\sigma}^{2}g_{2\sigma}^{2}t_{ij}$
is represented by open polygons.
Here,
we do not distinguish
holons and doublons.
Spins are also not distinguished in the diagram,
for simplicity.
\label{Dyson_2}}
\end{figure}

Here we construct approximated Green's functions for the Gaussian fluctuations of the bosons,
quasiparticles, and cofermions by using a set of the Dyson equations as is depicted in Fig.\ref{Dyson_2}:
Thick lines and thick wavy lines
stand for the Green's functions of the charge bosons
$
\mathcal{A}^{ab}(r,\tau)
=-\avrg{T\beta_{i}^{a}(\tau){\beta_{j}^{b}}^{\dagger}(0)},
$
and the spin bosons
$
\mathcal{C}^{ab}(r,\tau)
=-\avrg{T\phi_{i}^{a}(\tau){\phi_{j}^{b}}^{\dagger}(0)},
$
respectively, where $a,b=1,2$, $r=i-j$, $(\beta_{i}^{1},\beta_{i}^{2})=(\wdtn{e}{i},\wdtn{d}{i})$,
and $(\phi_{i}^{1},\phi_{i}^{2})=(\wdtn{p}{i\sigma},\wdtd{p}{i\overline{\sigma}})$.
Thick lines with arrows represent
the quasiparticles $\mathcal{G}_{\sigma}^{(f)}(r,\tau)$.
On the other hand, Thin lines and thin wavy lines represent
bare propagators of the charge bosons $\mathcal{A}_{0}^{ab}(r,\tau)$, the spin bosons $\mathcal{C}_{0}^{ab}(r,\tau)$, respectively,
determined by $\hat{\mathcal{L}}_{{\rm B}}^{(0)}$,
in which self-energy effects are not taken into account.
Thin lines with arrows stand for bare propagators of the quasiparticles
$\mathcal{G}_{0\sigma}^{(f)}(r,\tau)$ determined by
\eqsa{
	\hat{\mathcal{L}}_{0}=\sum_{ij}
	\hatd{f}{i\sigma}(\tau)
	[
	\hat{D}_{i}\delta_{ij}
	+
	\zeta_{0\sigma}t_{ij}
	]
	\hatn{f}{j\sigma}(\tau),
}
where $\zeta_{0\sigma}=g_{1\sigma}^{2}g_{2\sigma}^{2}
(\conp{p}{\sigma}\cond{e}+\cond{d}\conp{p}{\overline{\sigma}})^{2}$.
The Lagrangian $\hat{\mathcal{L}}_{0}$ is obtained by decoupling the fluctuating bosons
from the Lagrangian $\hat{\mathcal{L}}_{0}$ (Eq.(\ref{KR_Hubbard_2})).
Thick and thin dashed lines stand for the cofermions' propagators $\mathcal{F}^{ab}$ and bare propagators {$\mathcal{F}^{ab}_{0}=\delta_{a,b}/\epsilon$ $(\epsilon\rightarrow 0)$}, respectively.

In the set of Dyson equations (Fig.\ref{Dyson_2}),
we neglect the coupling between charge and spin bosons described by
propagators
such as
$\avrg{\wdtd{p}{i\sigma}\wdtn{e}{i}}$,
at the Gaussian level, since these coupling terms
are higher order contributions.
Below we explain that the coupling gives higher order contribution
with respect to hole-doping rate $x$, in proximity to Mott insulating states:
Since operators including both charge and spin such as $\wdtd{p}{i\sigma}\wdtn{e}{i}$
do not conserve the electric charge, propagators such as
$\avrg{\wdtd{p}{i\sigma}\wdtn{e}{i}}$ should vanish in the Mott insulating phase, where the charge
can not fluctuate.
Therefore,
the charge and spin excitations
are well separated in the Mott insulating phase.

When hole carriers are doped, $\wdtd{p}{i\sigma}\wdtn{e}{i}$ can have a
non zero expectation value, at most, scaled by condensate fraction of holons $\overline{e}_{0}$
which gives a rough estimate of the amplitude of charge fluctuations.
From a relation $\overline{e}_{0}^2\propto x$ held in the KR theory for the hole-doped case,
we obtain $\avrg{\wdtd{p}{i\sigma}\wdtn{e}{i}}\propto \sqrt{x}$.
Futhermore,
there is an additional constraint for the coupling terms such as $\avrg{\wdtd{p}{i\sigma}\wdtn{e}{i}}$:
they do not appear alone in calculations of physical quantities.
To conserve charge and spin on average,  $\avrg{\wdtd{p}{i\sigma}\wdtn{e}{i}}$ appears with $\avrg{\wdtd{e}{i}\wdtn{p}{i\sigma}}$ in pair, for example.
Therefore, the contribution of the coupling between charge and spin bosons to
physical quantities is scaled by $(\sqrt{x})^2$.
It concludes that the coupling between the charge and spin bosons gives contributions
as a higher order in terms of $x$ in physical quantities.

By solving the set of the Dyson equations,
we obtain the propagators for the quasiparticles and cofermions.
Here the bosonic degrees of freedom are taken into account in self-consistent fashion, 
through the cofermion self-energy
$\mbox{\boldmath$\Sigma$}^{({\rm cf})}_{\sigma}(r,\tau)$, and the amplitude
$\mbox{\boldmath$\Delta$}_{ij}$ of hybridization
between the quasiparticles and cofermions,
each of which
we detail below.

The Lagrangian for the cofermions is given by
{\eqsa{
	\hat{\mathcal{L}}_{{\rm cf}}
	=
	-
	\sum_{ij\sigma}
	\hat{\mbox{\boldmath$\Upsilon$}}_{i\sigma}^{\dagger}(\tau)
	\left[
	\mbox{\boldmath$\Sigma$}^{({\rm cf})}_{\sigma}(r,\tau)
	\right]
	\hat{\mbox{\boldmath$\Upsilon$}}_{j\sigma}(\tau),
}
where $\hat{\mbox{\boldmath$\Upsilon$}}_{i\sigma}^{\dagger}=(\hatd{\psi}{i\sigma}, \hatd{\chi}{i\sigma})$
is a vector notation for the cofermions, as is defined in the main article, and $r=i-j$.
The cofermion self-energy $\mbox{\boldmath$\Sigma$}^{({\rm cf})}_{\sigma}(r,\tau)$ is a 2$\times$2 symmetric matrix,
\eqsa{
	\mbox{\boldmath$\Sigma$}^{({\rm cf})}_{\sigma}
	=
	\left[
	\begin{array}{cc}
	\Sigma_{\sigma}^{11}&\Sigma'_{\sigma}\\
	\Sigma'_{\sigma}&\Sigma_{\sigma}^{22}\\
	\end{array}
	\right].
}
The hybridization between the quasiparticles and cofermions is described by
\eqsa{
	\hat{\mathcal{L}}_{{\rm hyb}}
	=
	\sum_{i,j,\sigma}
	[
	\hat{\mbox{\boldmath$\Upsilon$}}_{i\sigma}^{\dagger}(\tau)
	\mbox{\boldmath$\Delta$}_{ij}
	\hatn{f}{j\sigma}(\tau)
	+
	\hatd{f}{i\sigma}(\tau)
	\mbox{\boldmath$\Delta$}_{ij}^{T}
	\hat{\mbox{\boldmath$\Upsilon$}}_{j\sigma}(\tau)
	],
}
where $\mbox{\boldmath$\Delta$}_{ij}^{T}=(\Delta_{ij}^{(\psi)}, \Delta_{ij}^{(\chi)})$.
As a result,
the effective Lagrangian for the quasiparticles and cofermions, $\hat{\mathcal{L}}_{{\rm eff}}$, is 
given as
\eqsa{
\hat{\mathcal{L}}_{{\rm eff}}=\hat{\mathcal{L}}_{0}+\hat{\mathcal{L}}_{{\rm cf}}+\hat{\mathcal{L}}_{{\rm hyb}}.
}

When the charge gap is relatively small, $\Sigma_{\sigma}^{11}\simeq\Sigma_{\sigma}^{22}$
and $\Delta_{ij}^{(\psi)}\simeq\Delta_{ij}^{(\chi)}$
hold approximately. When the charge gap collapses, $\Sigma_{\sigma}^{11}=\Sigma_{\sigma}^{22}$
and $\Delta_{ij}^{(\psi)}=\Delta_{ij}^{(\chi)}$ hold exactly.}
In our results, we employ the approximate relations
$\Sigma=\Sigma_{\sigma}^{11}\simeq\Sigma_{\sigma}^{22}$ and
$\Delta_{ij}=\Delta_{ij}^{(\psi)}\simeq\Delta_{ij}^{(\chi)}$.
Then a cofermion mode $(\hatn{\psi}{k\sigma}+\hatn{\chi}{k\sigma})/\sqrt{2}$
hybridize with quasiparticles through the amplitude $\Delta (k)$, which is
depicted in Fig.\ref{Dyson_2} as closed (blue) circles, where $k$ is a momentum.
The inverse of cofermion propagator (namely, the cofermion self-energy)
for $(\hatn{\psi}{k\sigma}+\hatn{\chi}{k\sigma})/\sqrt{2}$
is given as
\eqsa{
\frac{-1}{2}[\Sigma_{\sigma}(k,i\varepsilon_{n})+\Sigma'_{\sigma}(k,i\varepsilon_{n})]
=
\gamma_{k}i\varepsilon_{n}-\alpha_{k}+O(\varepsilon_{n}^{2}),
}
where $\varepsilon_{n}$ is a fermionic Matsubara frequency.

Then,
the Green's function for the quasiparticles is
given as
\eqsa{
	&\mathcal{G}_{\sigma}^{(f)}(k,i\varepsilon_{n}\rightarrow \omega +i\delta)
	=
	G_{\sigma}^{(f)}(k,\omega+i\delta)
	\nn
	&\simeq
	\displaystyle
	\left[\omega+i\delta
	-\zeta_{0\sigma}\epsilon_{k}
	+\mu-
	\frac{
	\Delta(k)^{2}
	}{
	\gamma_{k}
	(\omega
	+i\delta)
	-\alpha_{k}
	}
	\right]^{-1},
}
where $\epsilon_{k}$ is the Fourier transformation of $t_{ij}$, and $\mu$
is the chemical potential.
Here we note that
the weights of the two quasiparticle bands split by the zero surface defined by $\omega=\alpha_{k}/\gamma_{k}$
are not the same in our theory.

In our calculations, we define the doping rate $x$ by using the quasiparticle Green's function as
\eqsa{
	1-x=\lim_{T\rightarrow 0+}\frac{T}{N_{s}}\sum_{k,i\varepsilon_{n},\sigma}\mathcal{G}_{\sigma}^{(f)}(k,i\varepsilon_{n}),
}
where $T$ stands for temperature and $N_{s}$ is the number of sites.

The Green's function for the electrons, instead of the quasiparticles,
is given as
\eqsa{
	\mathcal{G}_{ij\sigma}(\tau)
	&=&-\avrg{T \hatn{c}{i\sigma}(\tau)\hatd{c}{j\sigma}(0)}
	\nn
	&\simeq&-\avrg{T \hatd{z}{i\sigma}(\tau)\hatn{z}{j\sigma}(0)}
	\nn
	&\times&\avrg{T \hatn{f}{i\sigma}(\tau)\hatd{f}{j\sigma}(0)}
	\nn
	&=&\avrg{T \hatd{z}{i\sigma}(\tau)\hatn{z}{j\sigma}(0)}
	\mathcal{G}^{(f)}_{ij\sigma}(\tau),\label{GREEN_16}
}
where the bosonic and fermionic degrees of freedom are
decoupled, because the resultant action in our theory
does not contain the hybridization between bosons and fermions.
The quasiparticle Green's function is defined, as in
the previous sections, as
\eqsa{
	\mathcal{G}^{(f)}_{ij\sigma}(\tau)
	=-\avrg{T \hatn{f}{i\sigma}(\tau)\hatd{f}{j\sigma}(0)}.
}
The bosonic part in Eq.(\ref{GREEN_16}) is given by
\eqsa{
	\avrg{T \hatd{z}{i\sigma}(\tau)\hatn{z}{j\sigma}(0)}
	&\simeq& 
	g_{1\sigma}^{2}g_{2\sigma}^{2}
	\left\langle
	T [\hatd{\bvec{b}}{i}(\tau)\cdot \hatn{\bvec{p}}{i\sigma}(\tau)]
	\right.
	\nn
	&\times&
	\left.
	[\hatd{\bvec{p}}{j\sigma}(\tau)\cdot \hatn{\bvec{b}}{j}(\tau)]
	\right\rangle,
}
where we use vector notation as
$\hatd{\bvec{b}}{i}=(\hatd{e}{i},\hatn{d}{i})$, 
$\hatd{\bvec{p}}{i\sigma}=(\hatd{p}{i\sigma},\hatn{p}{i\overline{\sigma}})$.
Because
we adopt the boson dynamics in which
charge and spin bosons are decoupled,
this bosonic part of the Green's function is
rewritten as
\eqsa{
	&&\avrg{T \hatd{z}{i\sigma}(\tau)\hatn{z}{j\sigma}(0)}
	\nn
	&&\simeq 
	g_{1\sigma}^{2}g_{2\sigma}^{2}
	\left\langle
	T [\overline{\bvec{b}}_{0}\cdot \overline{\bvec{p}}_{0\sigma}^{T}]
	[\overline{\bvec{p}}_{0\sigma}\cdot \overline{\bvec{b}}_{0}^{T}]
	\right\rangle
	\nn
	&&+
	g_{1\sigma}^{2}g_{2\sigma}^{2}
	\left\langle
	T [\wdtd{\bvec{b}}{i}(\tau)\cdot \overline{\bvec{p}}_{0\sigma}^{T}]
	[\overline{\bvec{p}}_{0\sigma}\cdot \wdtn{\bvec{b}}{j}(\tau)]
	\right\rangle
	\nn
	&&+ 
	g_{1\sigma}^{2}g_{2\sigma}^{2}
	\left\langle
	T [\overline{\bvec{b}}_{0}\cdot \wdtn{\bvec{p}}{i\sigma}(\tau)]
	[\wdtd{\bvec{p}}{j\sigma}(\tau)\cdot \overline{\bvec{b}}_{0}^{T}]
	\right\rangle
	\nn
	&&+ 
	g_{1\sigma}^{2}g_{2\sigma}^{2}
	\left\langle
	T [\wdtd{\bvec{b}}{i}(\tau)\cdot \wdtn{\bvec{p}}{i\sigma}(\tau)]
	[\wdtd{\bvec{p}}{j\sigma}(\tau)\cdot \wdtn{\bvec{b}}{j}(\tau)]
	\right\rangle,\label{B_PART}
}
where
$\overline{\bvec{b}}_{0}=(\cond{e},\cond{d})$ and
$\overline{\bvec{p}}_{0\sigma}=(\conp{p}{\sigma},\conp{p}{\overline{\sigma}})$.
If we retain only the first and second lines of the right hand side of Eq.(\ref{B_PART}),
the electron Green's function is reduced to that already obtained in Ref.\onlinecite{Raimondi}. 
The contribution of the fourth line of the right hand side of Eq.(\ref{B_PART}) is
small compared with these from other lines
in Eq.(\ref{B_PART}), and we ignore the fourth term.
\subsection{{Supplementary result for quasiparticle density of states}}\label{A2}
\begin{figure}[t]
\begin{center}
\includegraphics[width=8.5cm]{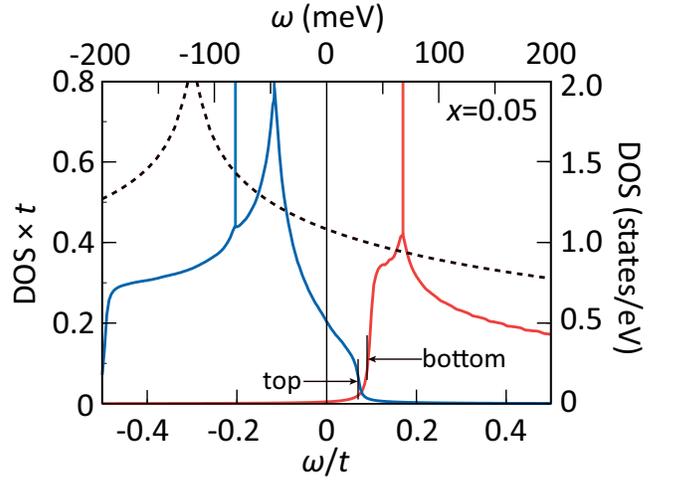}
\end{center}
\caption{
Quasiparticle density of states (DOS).
Solid curves show
our result for
DOS as function of $\omega$
for $x=0.05$.
Dashed curve shows DOS for the non-interacting case
with the same band parameter $t'=0.25t$ and the doping $x=0.05$.
\label{DOS_x005}}
\end{figure}
Here we show supplementary results for the $\omega$-dependence of
the quasiparticle density of states (DOS).
The $\omega$-dependence of DOS (solid curves in Fig.\ref{DOS_x005})
shows significant asymmetry around $\omega=0$
compared to the DOS for the non-interacting case
(the dashed curve in Fig.\ref{DOS_x005}).

This asymmetry of the DOS naturally explain
the asymmetric
tunneling spectra
with respect to the sign of the bias
observed in the scanning tunneling microscopy (STM)
measurements
of the hole-underdoped cuptrates
\cite{Renner98,Hanaguri04}.
Although in the STM measurements, the asymmetry is
observed up to several hundreds meV,
the present result offers a possible origin
of the asymmetry observed especially up to
100 meV.
Our result is in sharp contrast to the previous work
by Anderson and Ong\cite{Anderson67}, in which
the quasiparticle weights of the states for the added electrons
above $\mu$ and
the states for the removed electrons
below $\mu$ are different from each other.
They claimed that 
the quasiparticle weights inevitably show a step-like singularity
at the Fermi level
in the proximity to Mott insulators.
In our theory, such singularities at the Fermi level
are not needed for occurrence of the asymmetric DOS.
On the other hand, the recent study by Nieminen {\it et al}.,
shows that layers other than the CuO$_2$ layers in cuprates\cite{Nieminen09}
play a considerable role
on the tunneling spectra and cause asymmetric spectra.
Such effects of the layers other than the CuO$_2$ layers,
which are ignored
in our theory, will enhance the asymmetry of 
the tunneling spectra.

\end{document}